# ON THE RECURRENCE CHARACTER OF THE HADAMARD WALK IN THE PLANE

**CLEMENT AMPADU**

31 Carrolton Road
Boston, Massachusetts, 02132
USA
e-mail: drampadu@hotmail.com

### Abstract

We study return to the origin on $Z^2$ and investigate the stability properties of the Hadamard walk in the presence of bias.

## I. Introduction

The random walk which is popular [1,2] has been investigated extensively by many authors. The first random walks appeared in finance and biology [3,4] . The probability of the walker to return to its starting position on an infinite lattice, is often characterized by its Polya number [5]. If the Polya number is unity, the walk is said to be recurrent, otherwise the walk is said to be transient. The recurrent behavior of random walks has been studied in great detail for classical random walks in dependence on dimension and the topology of the lattice [6,7]. Other related studies on recurrence can be found in [8-19]. Recently in [20] the concept of Polya number was extended to quantum walks. Applying their definition of the Polya number to quantum walks on the line, it is shown in [21] that the recurrence character of quantum walks is more stable against bias. The range of parameters for which the biased quantum walk remains recurrent is determined. It is also found that there exists genuine biased quantum walks which are recurrent, which is the focus of our open problem in this paper in quantum wonderland.

The present paper focuses on the recurrent character for a one-parameter family of discrete-time quantum walk in the plane governed by the Hadamard operator. In Section II we describe the biased quantum walk in the plane and solve the time evolution equations with the help of the Fourier transformation. In Section III we use the method of stationary phase to perform asymptotic analysis of the probability at the origin in order to find the condition under which the biased quantum walk on the plane is recurrent. We also find that recurrence is related to the velocities of the peaks of the probability distribution. The explicit form of the velocities gives the same condition for recurrence when the method of stationary phase is employed. As for the method of stationary phase, the reader is referred to [22], it should also be noted that Nayak and Vishwanath in [23] have also given another description in the case of asymptotic expansions of integrals, in Appendix A of their paper. Section IV is devoted to the conclusion and an open problem concerning expressing the condition for recurrence in terms of the mean-value of the particles position in the plane.

**II. The Biased Quantum Walk in the Plane**

Consider the biased quantum walk in the plane where the particle has the possibility of moving to the left, right, up, or down. Assume the jump to the right or upward is of length $r$, respectively, and the jump to the left or downward is of unit length, respectively. The Hilbert space of the particle has the form of the tensor product $H = H_p \otimes H_c$ of the position space $H_p = span\{|x, y\rangle : x, y \in Z\}$, and the four dimensional coin space $H_C = span\{|R\rangle, |L\rangle, |U\rangle, |D\rangle\}$, where $R, L, U, D$ refers to the right, left, up, and downward chirality states of the particle. A single step of the quantum walk is given by $U = S(I_p \otimes C)$, where $I_p$ is the unit operator acting on the position space $H_p$, $S$ is the displacement operator which is defined by

$$S = \sum_{x,y=-\infty}^{\infty} \left( |x+r, y\rangle\langle x, y| \otimes |R\rangle\langle R| + |x-1, y\rangle\langle x, y| \otimes |L\rangle\langle L| + |x, y+r\rangle\langle x, y| \otimes |U\rangle\langle U| + |x, y-1\rangle\langle x, y| \otimes |D\rangle\langle D| \right)$$

and $C$ is an arbitrary unitary operator acting on the coin space $H_C$ which is applied to the coin

state before the displacement $S$ itself. In this paper we will take

$$C = \begin{pmatrix} p & \sqrt{p-p^2} & \sqrt{p-p^2} & 1-p \\ \sqrt{p-p^2} & -p & 1-p & -\sqrt{p-p^2} \\ \sqrt{p-p^2} & 1-p & -p & -\sqrt{p-p^2} \\ 1-p & -\sqrt{p-p^2} & -\sqrt{p-p^2} & -p \end{pmatrix}, \text{ where } p \in (0,1).$$

Let the initial state of the particle be given by $|\psi(0)\rangle = \sum_{x,y=-\infty}^{\infty} \sum_{i \in \{L,R,D,U\}} \psi_i(x,y,0)|x,y\rangle \otimes |i\rangle$. By induction on $t$, the state of the walker after $t$ steps can be defined in terms of the initial state by

$$|\psi(t)\rangle \equiv U^t |\psi(0)\rangle = \sum_{x,y=-\infty}^{\infty} \sum_{i \in \{L,R,D,U\}} \psi_i(x,y,t)|x,y\rangle \otimes |i\rangle.$$ The state of the particle is defined by

$$\psi(x,y,t) = \begin{bmatrix} \psi_R(x,y,t) \\ \psi_L(x,y,t) \\ \psi_U(x,y,t) \\ \psi_D(x,y,t) \end{bmatrix}, \text{ here } \psi_i(x,y,t) \text{ is the probability amplitude to find the particle at}$$

position $(x,y)$ at time $t$ with the initial coin state $|i\rangle$ for $i \in \{L,R,D,U\}$. The probability distribution generated by the quantum walk is given by $P(x,y,t) = \sum_{i \in \{L,R,D,U\}} |\psi_i(x,y,t)|^2 = \|\psi(x,y,t)\|$.

Put $C = C_R + C_L + C_U + C_D$, where $C_R = \begin{pmatrix} p & \sqrt{p-p^2} & \sqrt{p-p^2} & 1-p \\ 0 & 0 & 0 & 0 \\ 0 & 0 & 0 & 0 \\ 0 & 0 & 0 & 0 \end{pmatrix},$

$C_L = \begin{pmatrix} 0 & 0 & 0 & 0 \\ \sqrt{p-p^2} & -p & 1-p & -\sqrt{p-p^2} \\ 0 & 0 & 0 & 0 \\ 0 & 0 & 0 & 0 \end{pmatrix},$

$$C_U = \begin{pmatrix} 0 & 0 & 0 & 0 \\ 0 & 0 & 0 & 0 \\ \sqrt{p-p^2} & 1-p & -p & -\sqrt{p-p^2} \\ 0 & 0 & 0 & 0 \end{pmatrix}, \quad C_D = \begin{pmatrix} 0 & 0 & 0 & 0 \\ 0 & 0 & 0 & 0 \\ 0 & 0 & 0 & 0 \\ 1-p & -\sqrt{p-p^2} & -\sqrt{p-p^2} & -p \end{pmatrix}.$$

Then the time evolution equation $|\psi(t)\rangle \equiv U^t |\psi(0)\rangle = \sum_{x,y=-\infty}^{\infty} \sum_{i \in \{L,R,D,U\}} \psi_i(x,y,t) |x,y\rangle \otimes |i\rangle$, can be written as

$$\psi(x,y,t) = C_R \psi(x-r,y,t-1) + C_L \psi(x,y-1,t-1) + C_U \psi(x,y-r,t-1) + C_D \psi(x,y-1,t-1),$$

for the probability amplitude vectors $\psi(x,y,t)$. To simplify the time evolution we introduce the Fourier transform $\tilde{\psi}(k_x,k_y,t) = \sum_{x,y=-\infty}^{\infty} \psi(x,y,t) e^{i(k_x x + k_y y)}$ where $k_x, k_y \in [-\pi, \pi)$. Then in the Fourier domain we can write $\tilde{\psi}(k_x,k_y,t) = \tilde{U}(k_x,k_y) \tilde{\psi}(k_x,k_y,t-1)$, where

$$\tilde{U}(k_x,k_y) = \begin{pmatrix} e^{irk_x} & 0 & 0 & 0 \\ 0 & e^{-ik_x} & 0 & 0 \\ 0 & 0 & e^{irk_y} & 0 \\ 0 & 0 & 0 & e^{-irk_y} \end{pmatrix} \begin{pmatrix} p & \sqrt{p-p^2} & \sqrt{p-p^2} & 1-p \\ \sqrt{p-p^2} & -p & 1-p & -\sqrt{p-p^2} \\ \sqrt{p-p^2} & 1-p & -p & -\sqrt{p-p^2} \\ 1-p & -\sqrt{p-p^2} & -\sqrt{p-p^2} & -p \end{pmatrix}.$$

By induction on $t$ we can write $\tilde{\psi}(k_x,k_y,t) = \tilde{U}(k_x,k_y) \tilde{\psi}(k_x,k_y,t-1)$ in terms of $\tilde{\psi}(k_x,k_y,0)$ as

$\tilde{\psi}(k_x,k_y,t) = \tilde{U}^t(k_x,k_y) \tilde{\psi}(k_x,k_y,0)$, where $\tilde{\psi}(k_x,k_y,0)$ is the initial state in the Fourier domain.

We should remark from $\tilde{\psi}(k_x,k_y,t) = \sum_{x,y=-\infty}^{\infty} \psi(x,y,t) e^{i(k_x x + k_y y)}$, we can write

$$\tilde{\psi}(k_x,k_y,0) = \begin{bmatrix} \psi_R(0,0,0) \\ \psi_L(0,0,0) \\ \psi_U(0,0,0) \\ \psi_D(0,0,0) \end{bmatrix}. \text{ Let us denote } \tilde{\psi}(k_x,k_y,0) \text{ by } \psi = \begin{bmatrix} a \\ \sqrt{a-a^2} e^{i\varphi} \\ \sqrt{a-a^2} e^{i\varphi} \\ 1-ae^{i\varphi} \end{bmatrix}, \text{ where } a \in [0,1]$$

and $\varphi \in [0, 2\pi)$ are the parameterization parameters. If we denote the eigenvalues of $\tilde{U}(k_x,k_y)$

by $e^{iw_j(k_x,k_y)}$ and the corresponding eigenvectors by $|v_j(k_x,k_y)\rangle$, then in the Fourier domain the solution of the time evolution equation is given by

$$\tilde{\psi}(k_x,k_y,t) = \sum_{j=1}^{4} e^{iw_j(k_x,k_y)t} \langle v_j(k_x,k_y)|\psi\rangle |v_j(k_x,k_y)\rangle.$$ By the inverse Fourier transform we have

$$\psi(x,y,t) = \int_{-\pi}^{\pi}\int_{-\pi}^{\pi} \frac{dk_x}{2\pi}\frac{dk_y}{2\pi} \tilde{\psi}(k_x,k_y,t) e^{-i(k_x x + k_y y)} = \sum_{j=1}^{4}\int_{-\pi}^{\pi}\int_{-\pi}^{\pi} \frac{dk_x}{2\pi}\frac{dk_y}{2\pi} e^{i(w_j(k_x,k_y)t - k_x x - k_y y)} \langle v_j(k_x,k_y)|\psi\rangle |v_j(k_x,k_y)\rangle$$

Let $w_1\left(\dfrac{k_x \pm k_y}{2}\right) = \left(\dfrac{r-1}{2}\right)\left(\dfrac{k_x \pm k_y}{2}\right) + \arcsin\left(\sqrt{p}\sin\left(\left(\dfrac{r+1}{2}\right)\left(\dfrac{k_x \pm k_y}{2}\right)\right)\right)$ and

$w_2\left(\dfrac{k_x \pm k_y}{2}\right) = \left(\dfrac{r-1}{2}\right)\left(\dfrac{k_x \pm k_y}{2}\right) - \pi - \arcsin\left(\sqrt{p}\sin\left(\left(\dfrac{r+1}{2}\right)\left(\dfrac{k_x \pm k_y}{2}\right)\right)\right)$,

then it can be shown that the eigenvalues of $\tilde{U}(k_x,k_y)$ are given by

$$\lambda_1(k_x,k_y) = e^{i\left[w_1\left(\frac{k_x+k_y}{2}\right) + w_1\left(\frac{k_x-k_y}{2}\right)\right]}$$

$$\lambda_2(k_x,k_y) = e^{i\left[w_1\left(\frac{k_x+k_y}{2}\right) + w_2\left(\frac{k_x-k_y}{2}\right)\right]}$$

$$\lambda_3(k_x,k_y) = e^{i\left[w_1\left(\frac{k_x-k_y}{2}\right) + w_2\left(\frac{k_x+k_y}{2}\right)\right]}$$

$$\lambda_4(k_x,k_y) = e^{i\left[w_2\left(\frac{k_x+k_y}{2}\right) + w_2\left(\frac{k_x-k_y}{2}\right)\right]}$$

Concerning the eigenvectors of $\tilde{U}(k_x,k_y)$, let the appropriate normalization factors be defined as

$$n_1\left(\frac{(k_x \pm k_y)(r+1)}{4}\right) = 2 - 2\sqrt{p}\cos\left(\frac{(k_x \pm k_y)(r+1)}{4}\right) - \arcsin\left(\sqrt{p}\sin\left(\frac{(k_x \pm k_y)(r+1)}{4}\right)\right)$$

$$n_2\left(\frac{(k_x \pm k_y)(r+1)}{4}\right) = 2 + 2\sqrt{p}\cos\left(\frac{(k_x \pm k_y)(r+1)}{4} + \arcsin\left(\sqrt{p}\sin\left(\frac{(k_x \pm k_y)(r+1)}{4}\right)\right)\right), \text{ then,}$$

$$|v_1(k_x,k_y)\rangle = n_1\left(\frac{(k_x+k_y)(r+1)}{4}\right) n_1\left(\frac{(k_x-k_y)(r+1)}{4}\right) \begin{bmatrix} 1-p \\ -\sqrt{p(1-p)} + \sqrt{1-p}\, e^{i\left[w_1\left(\frac{k_x-k_y}{2}\right) - r\left(\frac{k_x-k_y}{2}\right)\right]} \\ -\sqrt{p(1-p)} + \sqrt{1-p}\, e^{i\left[w_1\left(\frac{k_x+k_y}{2}\right) - r\left(\frac{k_x+k_y}{2}\right)\right]} \\ p - \sqrt{p}e^{i\left[w_1\left(\frac{k_x-k_y}{2}\right) - r\left(\frac{k_x-k_y}{2}\right)\right]} - \sqrt{p}e^{i\left[w_1\left(\frac{k_x+k_y}{2}\right) - r\left(\frac{k_x+k_y}{2}\right)\right]} + e^{i\left[w_1\left(\frac{k_x+k_y}{2}\right) + w_1\left(\frac{k_x-k_y}{2}\right) - rk_x\right]} \end{bmatrix}$$

$$|v_2(k_x,k_y)\rangle = n_1\left(\frac{(k_x+k_y)(r+1)}{4}\right) n_2\left(\frac{(k_x-k_y)(r+1)}{4}\right) \begin{bmatrix} 1-p \\ -\sqrt{p(1-p)} + \sqrt{1-p}\, e^{i\left[w_2\left(\frac{k_x-k_y}{2}\right) - r\left(\frac{k_x-k_y}{2}\right)\right]} \\ -\sqrt{p(1-p)} + \sqrt{1-p}\, e^{i\left[w_1\left(\frac{k_x+k_y}{2}\right) - r\left(\frac{k_x+k_y}{2}\right)\right]} \\ p - \sqrt{p}e^{i\left[w_2\left(\frac{k_x-k_y}{2}\right) - r\left(\frac{k_x-k_y}{2}\right)\right]} - \sqrt{p}e^{i\left[w_1\left(\frac{k_x+k_y}{2}\right) - r\left(\frac{k_x+k_y}{2}\right)\right]} + e^{i\left[w_1\left(\frac{k_x+k_y}{2}\right) + w_2\left(\frac{k_x-k_y}{2}\right) - rk_x\right]} \end{bmatrix}$$

$$|v_3(k_x,k_y)\rangle = n_1\left(\frac{(k_x-k_y)(r+1)}{4}\right) n_2\left(\frac{(k_x+k_y)(r+1)}{4}\right) \begin{bmatrix} 1-p \\ -\sqrt{p(1-p)} + \sqrt{1-p}\, e^{i\left[w_1\left(\frac{k_x-k_y}{2}\right) - r\left(\frac{k_x-k_y}{2}\right)\right]} \\ -\sqrt{p(1-p)} + \sqrt{1-p}\, e^{i\left[w_2\left(\frac{k_x+k_y}{2}\right) - r\left(\frac{k_x+k_y}{2}\right)\right]} \\ p - \sqrt{p}e^{i\left[w_1\left(\frac{k_x-k_y}{2}\right) - r\left(\frac{k_x-k_y}{2}\right)\right]} - \sqrt{p}e^{i\left[w_2\left(\frac{k_x+k_y}{2}\right) - r\left(\frac{k_x+k_y}{2}\right)\right]} + e^{i\left[w_1\left(\frac{k_x-k_y}{2}\right) + w_2\left(\frac{k_x+k_y}{2}\right) - rk_x\right]} \end{bmatrix}$$

$$|v_4(k_x,k_y)\rangle = n_2\left(\frac{(k_x-k_y)(r+1)}{4}\right) n_2\left(\frac{(k_x+k_y)(r+1)}{4}\right) \begin{bmatrix} 1-p \\ -\sqrt{p(1-p)} + \sqrt{1-p}\, e^{i\left[w_2\left(\frac{k_x-k_y}{2}\right) - r\left(\frac{k_x-k_y}{2}\right)\right]} \\ -\sqrt{p(1-p)} + \sqrt{1-p}\, e^{i\left[w_2\left(\frac{k_x+k_y}{2}\right) - r\left(\frac{k_x+k_y}{2}\right)\right]} \\ p - \sqrt{p}e^{i\left[w_2\left(\frac{k_x-k_y}{2}\right) - r\left(\frac{k_x-k_y}{2}\right)\right]} - \sqrt{p}e^{i\left[w_2\left(\frac{k_x+k_y}{2}\right) - r\left(\frac{k_x+k_y}{2}\right)\right]} + e^{i\left[w_2\left(\frac{k_x+k_y}{2}\right) + w_2\left(\frac{k_x-k_y}{2}\right) - rk_x\right]} \end{bmatrix}$$

## III. Recurrence Nature of the Biased Quantum Walk

The determination of the recurrent nature of the quantum walk is based on the method of stationary phase [22]. This involves analysis of the asymptotic behavior of the probability at the origin. We should remark that from

$$\psi(x,y,t) = \int_{-\pi-\pi}^{\pi}\int^{\pi}\frac{dk_x}{2\pi}\frac{dk_y}{2\pi}\tilde{\psi}(k_x,k_y,t)e^{-i(k_x x+k_y y)} = \sum_{j=1}^{4}\int_{-\pi-\pi}^{\pi}\int^{\pi}\frac{dk_x}{2\pi}\frac{dk_y}{2\pi}e^{i(w_j(k_x,k_y)t-k_x x-k_y y)}\langle v_j(k_x,k_y)|\psi\rangle|v_j(k_x,k_y)\rangle$$

the amplitude at the origin is given by

$$\psi(0,0,t) = \sum_{j=1}^{4}\int_{-\pi-\pi}^{\pi}\int^{\pi}\frac{dk_x}{2\pi}\frac{dk_y}{2\pi}e^{i(w_j(k_x,k_y)t)}\langle v_j(k_x,k_y)|\psi\rangle|v_j(k_x,k_y)\rangle$$ which allows us to find the

asymptotics of the probability at the origin. We should note that the important contributions to the amplitude at the origin come from the stationary points of $w_j(k_x,k_y)$. Let us now consider the derivatives of the phases. We find it convenient to write

$$w_1(k_x,k_y) = w_1\left(\frac{k_x+k_y}{2}\right) + w_1\left(\frac{k_x-k_y}{2}\right)$$

$$w_2(k_x,k_y) = w_1\left(\frac{k_x+k_y}{2}\right) + w_2\left(\frac{k_x-k_y}{2}\right)$$

$$w_3(k_x,k_y) = w_1\left(\frac{k_x-k_y}{2}\right) + w_2\left(\frac{k_x+k_y}{2}\right)$$

$$w_4(k_x,k_y) = w_2\left(\frac{k_x+k_y}{2}\right) + w_2\left(\frac{k_x-k_y}{2}\right)$$

It follows from partial differentiation that we have the following

$$\frac{\partial w_1}{\partial k_x} = \frac{r-1}{2} + \frac{\sqrt{p}\cos\left(\left(\frac{r+1}{2}\right)\left(\frac{k_x+k_y}{2}\right)\right)\left(\frac{r+1}{2}\right)\left(\frac{1}{2}\right)}{\sqrt{1-p\sin^2\left(\left(\frac{r+1}{2}\right)\left(\frac{k_x+k_y}{2}\right)\right)}} + \frac{\sqrt{p}\cos\left(\left(\frac{r+1}{2}\right)\left(\frac{k_x-k_y}{2}\right)\right)\left(\frac{r+1}{2}\right)\left(\frac{1}{2}\right)}{\sqrt{1-p\sin^2\left(\left(\frac{r+1}{2}\right)\left(\frac{k_x-k_y}{2}\right)\right)}}$$

$$\frac{\partial w_1}{\partial k_y} = \frac{\sqrt{p}\cos\left(\left(\frac{r+1}{2}\right)\left(\frac{k_x+k_y}{2}\right)\right)\left(\frac{r+1}{2}\right)\left(\frac{1}{2}\right)}{\sqrt{1-p\sin^2\left(\left(\frac{r+1}{2}\right)\left(\frac{k_x+k_y}{2}\right)\right)}} - \frac{\sqrt{p}\cos\left(\left(\frac{r+1}{2}\right)\left(\frac{k_x-k_y}{2}\right)\right)\left(\frac{r+1}{2}\right)\left(\frac{1}{2}\right)}{\sqrt{1-p\sin^2\left(\left(\frac{r+1}{2}\right)\left(\frac{k_x-k_y}{2}\right)\right)}}$$

$$\frac{\partial w_2}{\partial k_x} = \frac{r-1}{2} + \frac{\sqrt{p}\cos\left(\left(\frac{r+1}{2}\right)\left(\frac{k_x+k_y}{2}\right)\right)\left(\frac{r+1}{2}\right)\left(\frac{1}{2}\right)}{\sqrt{1-p\sin^2\left(\left(\frac{r+1}{2}\right)\left(\frac{k_x+k_y}{2}\right)\right)}} - \frac{\sqrt{p}\cos\left(\left(\frac{r+1}{2}\right)\left(\frac{k_x-k_y}{2}\right)\right)\left(\frac{r+1}{2}\right)\left(\frac{1}{2}\right)}{\sqrt{1-p\sin^2\left(\left(\frac{r+1}{2}\right)\left(\frac{k_x-k_y}{2}\right)\right)}}$$

$$\frac{\partial w_2}{\partial k_y} = \frac{\sqrt{p}\cos\left(\left(\frac{r+1}{2}\right)\left(\frac{k_x+k_y}{2}\right)\right)\left(\frac{r+1}{2}\right)\left(\frac{1}{2}\right)}{\sqrt{1-p\sin^2\left(\left(\frac{r+1}{2}\right)\left(\frac{k_x+k_y}{2}\right)\right)}} + \frac{\sqrt{p}\cos\left(\left(\frac{r+1}{2}\right)\left(\frac{k_x-k_y}{2}\right)\right)\left(\frac{r+1}{2}\right)\left(\frac{1}{2}\right)}{\sqrt{1-p\sin^2\left(\left(\frac{r+1}{2}\right)\left(\frac{k_x-k_y}{2}\right)\right)}}$$

$$\frac{\partial w_3}{\partial k_x} = \frac{r-1}{2} + \frac{\sqrt{p}\cos\left(\left(\frac{r+1}{2}\right)\left(\frac{k_x-k_y}{2}\right)\right)\left(\frac{r+1}{2}\right)\left(\frac{1}{2}\right)}{\sqrt{1-p\sin^2\left(\left(\frac{r+1}{2}\right)\left(\frac{k_x-k_y}{2}\right)\right)}} - \frac{\sqrt{p}\cos\left(\left(\frac{r+1}{2}\right)\left(\frac{k_x+k_y}{2}\right)\right)\left(\frac{r+1}{2}\right)\left(\frac{1}{2}\right)}{\sqrt{1-p\sin^2\left(\left(\frac{r+1}{2}\right)\left(\frac{k_x+k_y}{2}\right)\right)}}$$

$$\frac{\partial w_3}{\partial k_y} = -\frac{\sqrt{p}\cos\left(\left(\frac{r+1}{2}\right)\left(\frac{k_x-k_y}{2}\right)\right)\left(\frac{r+1}{2}\right)\left(\frac{1}{2}\right)}{\sqrt{1-p\sin^2\left(\left(\frac{r+1}{2}\right)\left(\frac{k_x-k_y}{2}\right)\right)}} - \frac{\sqrt{p}\cos\left(\left(\frac{r+1}{2}\right)\left(\frac{k_x+k_y}{2}\right)\right)\left(\frac{r+1}{2}\right)\left(\frac{1}{2}\right)}{\sqrt{1-p\sin^2\left(\left(\frac{r+1}{2}\right)\left(\frac{k_x+k_y}{2}\right)\right)}}$$

$$\frac{\partial w_4}{\partial k_x} = \frac{r-1}{2} - \frac{\sqrt{p}\cos\left(\left(\frac{r+1}{2}\right)\left(\frac{k_x+k_y}{2}\right)\right)\left(\frac{r+1}{2}\right)\left(\frac{1}{2}\right)}{\sqrt{1-p\sin^2\left(\left(\frac{r+1}{2}\right)\left(\frac{k_x+k_y}{2}\right)\right)}} - \frac{\sqrt{p}\cos\left(\left(\frac{r+1}{2}\right)\left(\frac{k_x-k_y}{2}\right)\right)\left(\frac{r+1}{2}\right)\left(\frac{1}{2}\right)}{\sqrt{1-p\sin^2\left(\left(\frac{r+1}{2}\right)\left(\frac{k_x-k_y}{2}\right)\right)}}$$

$$\frac{\partial w_4}{\partial k_y} = -\frac{\sqrt{p}\cos\left(\left(\frac{r+1}{2}\right)\left(\frac{k_x+k_y}{2}\right)\right)\left(\frac{r+1}{2}\right)\left(\frac{1}{2}\right)}{\sqrt{1-p\sin^2\left(\left(\frac{r+1}{2}\right)\left(\frac{k_x+k_y}{2}\right)\right)}} + \frac{\sqrt{p}\cos\left(\left(\frac{r+1}{2}\right)\left(\frac{k_x-k_y}{2}\right)\right)\left(\frac{r+1}{2}\right)\left(\frac{1}{2}\right)}{\sqrt{1-p\sin^2\left(\left(\frac{r+1}{2}\right)\left(\frac{k_x-k_y}{2}\right)\right)}}$$

By the method of stationary phase the amplitudes will decay at a rate inversely proportional to the square root of $t$, if at least one of the phases has a vanishing derivative inside the integration domain. It therefore becomes necessary to solve the equations

$$\frac{\partial w_1}{\partial k_x}=0, \ \frac{\partial w_1}{\partial k_y}=0, \ \frac{\partial w_2}{\partial k_x}=0, \ \frac{\partial w_2}{\partial k_y}=0, \ \frac{\partial w_3}{\partial k_x}=0, \ \frac{\partial w_3}{\partial k_y}=0, \ \frac{\partial w_4}{\partial k_x}=0, \ \frac{\partial w_4}{\partial k_y}=0, \text{ to find the saddle}$$

points. Since the phases are functions of two variables we will use the following.

**Theorem (Second Derivative Test):** Let $w_i(k_x, k_y)$ be continuous in a disk with center $(a,b)$, and suppose that $\dfrac{\partial w_i(a,b)}{\partial k_x} = 0$ and $\dfrac{\partial w_i(a,b)}{\partial k_y} = 0$ for $i = 1,2,3,4$. Let

$D(a,b) = \dfrac{\partial^2 w_i(a,b)}{\partial k_x^2}\dfrac{\partial^2 w_i(a,b)}{\partial k_y^2} - \left[\dfrac{\partial^2 w_i(a,b)}{\partial k_y \partial k_x}\right]^2$. If $D(a,b) < 0$, then $w_i(a,b)$ is not a local

extremum, in particular $(a,b)$ is a saddle point of $w_i$.

If we consider the first phase, it can be shown that $\dfrac{\partial w_1(k_x,k_y)}{\partial k_x} = 0$, and $\dfrac{\partial w_1(k_x,k_y)}{\partial k_y} = 0$ gives the

following

$\dfrac{k_{x_0}+k_{y_0}}{2}, \dfrac{k_{x_0}-k_{y_0}}{2} = \dfrac{2}{r+1}\arcsin\left(\pm\sqrt{\dfrac{p(r+1)^2-(r-1)^2}{p(r+1)^2-p(r-1)^2}}\right)$. Thus $w_1$ gives the possible saddle point as $(k_{x_0}, k_{y_0}) = \left(\dfrac{4}{r+1}\arcsin\left(\pm\sqrt{\dfrac{p(r+1)^2-(r-1)^2}{p(r+1)^2-p(r-1)^2}}\right), 0\right)$. Now if we consider the second phase, it can be shown that $\dfrac{\partial w_2(k_x,k_y)}{\partial k_x}=0$ and $\dfrac{\partial w_2(k_x,k_y)}{\partial k_y}=0$ gives the following

$$\dfrac{k_{x_0}+k_{y_0}}{2},\dfrac{k_{x_0}-k_{y_0}}{2}=\dfrac{2}{r+1}\arcsin\left(\pm\sqrt{\dfrac{p(r+1)^2-(r-1)^2}{p(r+1)^2-p(r-1)^2}}\right).$$ Thus we also see that $w_2$ gives the

possible saddle points as $(k_{x_0},k_{y_0})=\left(\dfrac{4}{r+1}\arcsin\left(\pm\sqrt{\dfrac{p(r+1)^2-(r-1)^2}{p(r+1)^2-p(r-1)^2}}\right),0\right)$. With respect to the third phase, $\dfrac{\partial w_3(k_x,k_y)}{\partial k_x}=0$ and $\dfrac{\partial w_3(k_x,k_y)}{\partial k_y}=0$ also

$$\dfrac{k_{x_0}+k_{y_0}}{2},\dfrac{k_{x_0}-k_{y_0}}{2}=\dfrac{2}{r+1}\arcsin\left(\pm\sqrt{\dfrac{p(r+1)^2-(r-1)^2}{p(r+1)^2-p(r-1)^2}}\right).$$ Thus we also see that $w_3$ gives the

possible saddle point as $(k_{x_0},k_{y_0})=\left(\dfrac{4}{r+1}\arcsin\left(\pm\sqrt{\dfrac{p(r+1)^2-(r-1)^2}{p(r+1)^2-p(r-1)^2}}\right),0\right)$. Finally from

$\dfrac{\partial w_4(k_x,k_y)}{\partial k_x}=0$ and $\dfrac{\partial w_4(k_x,k_y)}{\partial k_y}=0$, we can conclude that $w_4$ gives the possible saddle points as

$(k_{x_0},k_{y_0})=\left(\dfrac{4}{r+1}\arcsin\left(\pm\sqrt{\dfrac{p(r+1)^2-(r-1)^2}{p(r+1)^2-p(r-1)^2}}\right),0\right)$. Since all the phases give

$(k_{x_0},k_{y_0})=\left(\dfrac{4}{r+1}\arcsin\left(\pm\sqrt{\dfrac{p(r+1)^2-(r-1)^2}{p(r+1)^2-p(r-1)^2}}\right),0\right)$ as the only saddle points, it is not necessary

to check the Theorem above, in particular, the coordinates of the saddle points are real-valued only if

$\dfrac{p(r+1)^2-(r-1)^2}{p(r+1)^2-p(r-1)^2}\leq 1$, from which it follows we have the following.

**CONDITION FOR RECURRENCE:** $p\leq 1$

It should be noted that the condition on the parameter of the coin $p$ to be recurrent is independent of the size of the step to the right or the upward direction in the plane, namely $r$.

We now give an alternate condition for recurrence of the biased quantum walk in the plane. Consider the peaks of the probability distribution generated by the quantum walk. If the walk is recurrent then the origin of the walk has to remain between the peaks at all times. In particular let $v_L$, $v_R$, $v_U$, $v_D$ be the velocities of the left, right, upward, and, downward peaks, respectively, then we can make the following.

**CLAIM:** *The biased quantum walk in the plane is recurrent if and only if $v_L < 0$, $v_R > 0$, $v_U > 0$, and $v_D < 0$.*

**Proof of Claim:** We begin by noticing that we can write

$$\psi(x,y,t) = \sum_{j=1}^{4} \int_{-\pi}^{\pi}\int_{-\pi}^{\pi} \frac{dk_x}{2\pi}\frac{dk_y}{2\pi} e^{i(w_j(k_x,k_y)t - k_x x - k_y y)} \langle v_j(k_x,k_y)|\psi\rangle |v_j(k_x,k_y)\rangle \text{ as}$$

$$\psi(x,y,t) = \sum_{j=1}^{4} \int_{-\pi}^{\pi}\int_{-\pi}^{\pi} \frac{dk_x}{2\pi}\frac{dk_y}{2\pi} e^{i(w_j(k_x,k_y) - k_x \alpha_x - k_y \alpha_y)t} \langle v_j(k_x,k_y)|\psi\rangle |v_j(k_x,k_y)\rangle, \text{ where we have introduced}$$

$\alpha_x = \frac{x}{t}$, and $\alpha_y = \frac{y}{t}$. Now consider the modified phase $\tilde{w}_j(k_x, k_y) = w_j(k_x, k_y) - \alpha_x k_x - \alpha_y k_y$, the peak will occur at $(x_0, y_0)$ where both the first and second derivatives of $\tilde{w}_j$ vanishes, thus the components of the velocities are thus $\alpha_{x_0} = \frac{x_0}{t}$ and $\alpha_{y_0} = \frac{y_0}{t}$ in the $x-$ and $y-$ directions of the plane respectively. Now computing the first and second derivatives of $\tilde{w}_j$. For the first derivatives it is easily seen that for $j = 1,2,3,4$ we have

$$\frac{\partial \tilde{w}_j(k_x,k_y)}{\partial k_x} = \frac{\partial w_j(k_x,k_y)}{\partial k_x} - \alpha_x, \text{ where } \frac{\partial w_j(k_x,k_y)}{\partial k_x} \text{ are as given above, and}$$

$$\frac{\partial \tilde{w}_j(k_x,k_y)}{\partial k_y} = \frac{\partial w_j(k_x,k_y)}{\partial k_y} - \alpha_y, \text{ where } \frac{\partial w_j(k_x,k_y)}{\partial k_y} \text{ are as given above.}$$

As for the second derivatives for those not involving the mixed partials we have

$$\frac{\partial^2 \tilde{w}_j(k_x,k_y)}{\partial k_x^2} = \frac{\partial^2 w_j(k_x,k_y)}{\partial k_x^2}, \quad \frac{\partial^2 \tilde{w}_j(k_x,k_y)}{\partial k_y^2} = \frac{\partial^2 w_j(k_x,k_y)}{\partial k_y^2}, \text{ explicitly for } j=1,2,3,4 \text{ these}$$

become

$$\frac{\partial^2 \tilde{w}_1(k_x,k_y)}{\partial k_x^2} = \frac{\partial f}{\partial k_x} + \frac{\partial g}{\partial k_x}, \quad \frac{\partial^2 \tilde{w}_1(k_x,k_y)}{\partial k_y^2} = \frac{\partial f}{\partial k_y} - \frac{\partial g}{\partial k_y}$$

$$\frac{\partial^2 \tilde{w}_2(k_x,k_y)}{\partial k_x^2} = \frac{\partial f}{\partial k_x} - \frac{\partial g}{\partial k_x}, \quad \frac{\partial^2 \tilde{w}_2(k_x,k_y)}{\partial k_y^2} = \frac{\partial f}{\partial k_y} + \frac{\partial g}{\partial k_y}$$

$$\frac{\partial^2 \tilde{w}_3(k_x,k_y)}{\partial k_x^2} = \frac{\partial g}{\partial k_x} - \frac{\partial f}{\partial k_x}, \quad \frac{\partial^2 \tilde{w}_3(k_x,k_y)}{\partial k_y^2} = -\frac{\partial g}{\partial k_y} - \frac{\partial f}{\partial k_y}$$

$$\frac{\partial^2 \tilde{w}_4(k_x,k_y)}{\partial k_x^2} = -\frac{\partial f}{\partial k_x} - \frac{\partial g}{\partial k_x}, \quad \frac{\partial^2 \tilde{w}_4(k_x,k_y)}{\partial k_y^2} = -\frac{\partial f}{\partial k_y} + \frac{\partial g}{\partial k_y}$$

where

$$\frac{\partial f}{\partial k_x} = \frac{-\sqrt{p}\sin\left(\left(\frac{r+1}{2}\right)\left(\frac{k_x+k_y}{2}\right)\right)\frac{(r+1)^2}{16}}{\sqrt{1-p\sin^2\left(\left(\frac{r+1}{2}\right)\left(\frac{k_x+k_y}{2}\right)\right)}} - \frac{p\sqrt{p}\cos^2\left(\left(\frac{r+1}{2}\right)\left(\frac{k_x+k_y}{2}\right)\right)\sin\left(\left(\frac{r+1}{2}\right)\left(\frac{k_x+k_y}{2}\right)\right)\frac{(r+1)^2}{16}}{\left(1-p\sin^2\left(\left(\frac{r+1}{2}\right)\left(\frac{k_x+k_y}{2}\right)\right)\right)^{\frac{3}{2}}}$$

$$\frac{\partial f}{\partial k_y} = \frac{\sqrt{p}\sin\left(\left(\frac{r+1}{2}\right)\left(\frac{k_x+k_y}{2}\right)\right)\frac{(r+1)^2}{16}}{\sqrt{1-p\sin^2\left(\left(\frac{r+1}{2}\right)\left(\frac{k_x+k_y}{2}\right)\right)}} + \frac{p\sqrt{p}\cos^2\left(\left(\frac{r+1}{2}\right)\left(\frac{k_x+k_y}{2}\right)\right)\sin\left(\left(\frac{r+1}{2}\right)\left(\frac{k_x+k_y}{2}\right)\right)\frac{(r+1)^2}{16}}{\left(1-p\sin^2\left(\left(\frac{r+1}{2}\right)\left(\frac{k_x+k_y}{2}\right)\right)\right)^{\frac{3}{2}}}$$

$$\frac{\partial g}{\partial k_x} = \frac{-\sqrt{p}\sin\left(\left(\frac{r+1}{2}\right)\left(\frac{k_x-k_y}{2}\right)\right)\frac{(r+1)^2}{16}}{\sqrt{1-p\sin^2\left(\left(\frac{r+1}{2}\right)\left(\frac{k_x-k_y}{2}\right)\right)}} + \frac{p\sqrt{p}\cos^2\left(\left(\frac{r+1}{2}\right)\left(\frac{k_x-k_y}{2}\right)\right)\sin\left(\left(\frac{r+1}{2}\right)\left(\frac{k_x-k_y}{2}\right)\right)\frac{(r+1)^2}{16}}{\left(1-p\sin^2\left(\left(\frac{r+1}{2}\right)\left(\frac{k_x-k_y}{2}\right)\right)\right)^{\frac{3}{2}}}$$

$$\frac{\partial g}{\partial k_y} = \frac{\sqrt{p}\sin\left(\left(\frac{r+1}{2}\right)\left(\frac{k_x-k_y}{2}\right)\right)\frac{(r+1)^2}{16}}{\sqrt{1-p\sin^2\left(\left(\frac{r+1}{2}\right)\left(\frac{k_x-k_y}{2}\right)\right)}} - \frac{p\sqrt{p}\cos^2\left(\left(\frac{r+1}{2}\right)\left(\frac{k_x-k_y}{2}\right)\right)\sin\left(\left(\frac{r+1}{2}\right)\left(\frac{k_x-k_y}{2}\right)\right)\frac{(r+1)^2}{16}}{\left(1-p\sin^2\left(\left(\frac{r+1}{2}\right)\left(\frac{k_x-k_y}{2}\right)\right)\right)^{\frac{3}{2}}}$$

Now for the second derivatives involving the mixed partials we have the following

$$\frac{\partial^2 \tilde{w}_j(k_x,k_y)}{\partial k_y \partial k_x} = \frac{\partial^2 w_j(k_x,k_y)}{\partial k_y \partial k_x}, \quad \frac{\partial^2 \tilde{w}_j(k_x,k_y)}{\partial k_x \partial k_y} = \frac{\partial^2 w_j(k_x,k_y)}{\partial k_x \partial k_y}.$$ Explicitly, for $j=1,2,3,4$, upon using

the expressions for $\frac{\partial f}{\partial k_x}$, $\frac{\partial f}{\partial k_y}$, $\frac{\partial g}{\partial k_x}$, and $\frac{\partial g}{\partial k_y}$, we get the following

$$\frac{\partial^2 \tilde{w}_1(k_x,k_y)}{\partial k_y \partial k_x} = \frac{\partial^2 w_1(k_x,k_y)}{\partial k_y \partial k_x} = \frac{\partial f}{\partial k_y} + \frac{\partial g}{\partial k_y}$$

$$\frac{\partial^2 \tilde{w}_1(k_x,k_y)}{\partial k_x \partial k_y} = \frac{\partial^2 w_1(k_x,k_y)}{\partial k_x \partial k_y} = \frac{\partial f}{\partial k_x} - \frac{\partial g}{\partial k_x}$$

$$\frac{\partial^2 \tilde{w}_2(k_x,k_y)}{\partial k_y \partial k_x} = \frac{\partial^2 w_2(k_x,k_y)}{\partial k_y \partial k_x} = \frac{\partial f}{\partial k_y} - \frac{\partial g}{\partial k_y}$$

$$\frac{\partial^2 \tilde{w}_2(k_x,k_y)}{\partial k_x \partial k_y} = \frac{\partial^2 w_2(k_x,k_y)}{\partial k_x \partial k_y} = \frac{\partial f}{\partial k_x} + \frac{\partial g}{\partial k_x}$$

$$\frac{\partial^2 \tilde{w}_3(k_x,k_y)}{\partial k_y \partial k_x} = \frac{\partial^2 w_3(k_x,k_y)}{\partial k_y \partial k_x} = \frac{\partial g}{\partial k_y} - \frac{\partial f}{\partial k_y}$$

$$\frac{\partial^2 \tilde{w}_3(k_x,k_y)}{\partial k_x \partial k_y} = \frac{\partial^2 w_3(k_x,k_y)}{\partial k_x \partial k_y} = -\frac{\partial g}{\partial k_x} - \frac{\partial f}{\partial k_x}$$

$$\frac{\partial^2 \tilde{w}_4(k_x,k_y)}{\partial k_y \partial k_x} = \frac{\partial^2 w_4(k_x,k_y)}{\partial k_y \partial k_x} = -\frac{\partial g}{\partial k_y} - \frac{\partial f}{\partial k_y}$$

$$\frac{\partial^2 \tilde{w}_4(k_x,k_y)}{\partial k_x \partial k_y} = \frac{\partial^2 w_4(k_x,k_y)}{\partial k_x \partial k_y} = -\frac{\partial f}{\partial k_x} + \frac{\partial g}{\partial k_x}$$

Since the second derivatives are independent of $\alpha_x$ and $\alpha_y$ we can deduce upon setting them equal to zero, that it is necessary only to solve the following equations $\frac{\partial f}{\partial k_x} = 0$, $\frac{\partial f}{\partial k_y} = 0$, $\frac{\partial g}{\partial k_x} = 0$ and $\frac{\partial g}{\partial k_y} = 0$, in order to find the solutions. We see that in the case of $\frac{\partial g}{\partial k_x} = 0$ and $\frac{\partial g}{\partial k_y} = 0$, $\frac{k_{x_0} - k_{y_0}}{2}$ can assume any *value,* however, from $\frac{\partial f}{\partial k_x} = 0$, $\frac{\partial g}{\partial k_x} = 0$, we see that

$$\frac{k_{x_0} + k_{y_0}}{2} = \frac{2}{r+1} \arcsin\left(\pm\sqrt{\frac{1+p}{2p}}\right),$$ without loss of generality, we will set

$$\frac{k_{x_0} - k_{y_0}}{2} = \frac{2}{r+1} \arcsin\left(\pm\sqrt{\frac{1+p}{2p}}\right).$$ Now to determine the velocities we set the values into

$$\frac{\partial \tilde{w}_j(k_x, k_y)}{\partial k_x} = \frac{\partial w_j(k_x, k_y)}{\partial k_x} - \alpha_x \text{ and } \frac{\partial \tilde{w}_j(k_x, k_y)}{\partial k_y} = \frac{\partial w_j(k_x, k_y)}{\partial k_y} - \alpha_y \text{ for . We get the velocities}$$

$$v_{R_x} = \frac{r-1}{2} + \sqrt{p}\left(\frac{r+1}{2}\right), v_{R_y} = 0, v_{L_x} = \frac{r-1}{2}, v_{L_y} = \sqrt{p}\left(\frac{r+1}{2}\right), v_{D_x} = \frac{r-1}{2}, v_{D_y} = -\sqrt{p}\left(\frac{r+1}{2}\right),$$

and $v_{U_x} = \frac{r-1}{2} - \sqrt{p}\left(\frac{r+1}{2}\right)$, $v_{U_y} = 0$. Using the velocities in conjunction with the condition we obtained for recurrence earlier gives the desired result.

### IV. Concluding Remarks and Open Problem

In this paper we have obtained the condition for recurrence of the biased Hadamard walk in the plane using two different approaches. The open problem is an attempt to give another condition for recurrence in terms of the mean value of the biased quantum walk. In particular we make the following.

**Conjecture**: *The biased quantum walk in the plane is recurrent if and only if the mean value of the position vanishes*.

To show the truth or falsity of the conjecture it might be necessary to answer the following.

*Does there exist biased quantum walks on the plane which are recurrent but cannot produce probability distributions with zero mean value?*

Obviously, the problem becomes solvable as the mean value can be obtained by the weak limit theorem [24].